\documentclass[prd,nofootinbib,aps,superscriptaddress,preprintnumbers,11pt]{revtex4}
\usepackage[dvips]{graphicx}

\usepackage{amsmath,amssymb}
\usepackage{bm}
\usepackage{times}
\usepackage{epsfig}
\usepackage{verbatim}
\usepackage{psfrag}
\usepackage{bm}

\newcommand{\be}{\begin{equation}}
\newcommand{\ee}{\end{equation}}
\newcommand{\bea}{\begin{eqnarray}}
\newcommand{\eea}{\end{eqnarray}}
\newcommand{\ba}{\begin{array}}
\newcommand{\ea}{\end{array}}

\newcommand{\cpv}{{\textrm{\fontsize{6.5}{11}\selectfont CPV}}}

\begin{document}
\preprint{NPAC-09-05}
\vfill
\vspace{7.0cm}
\title{\Large Lepton-mediated electroweak baryogenesis}
\vspace{4.0cm}

\author{Daniel J. H. Chung}
\email{danielchung@wisc.edu}
\affiliation{University of Wisconsin, Madison, WI, 53706-1390}
\author{Bjorn Garbrecht}
\email{bjorn@physics.wisc.edu}
\affiliation{University of Wisconsin, Madison, WI, 53706-1390}
\author{Michael J. Ramsey-Musolf}
\email{mjrm@physics.wisc.edu}
\affiliation{University of Wisconsin, Madison, WI, 53706-1390}
\affiliation{California Institute of Technology, Pasadena, CA 91125}
\author{Sean Tulin} 
\email{tulin@caltech.edu}
\affiliation{California Institute of Technology, Pasadena, CA 91125}

\date{\today}
\begin{abstract}
We investigate the impact of the tau and bottom Yukawa couplings on the transport dynamics for electroweak baryogenesis in
supersymmetric extensions of the Standard Model.  
Although it has
generally been assumed in the literature that all Yukawa interactions except those involving the top quark are
negligible, we find that the tau and bottom Yukawa
interaction rates are too fast to be neglected.  
We identify an illustrative
``lepton-mediated electroweak baryogenesis'' scenario in which the baryon
asymmetry is induced mainly through the presence of a left-handed {\it
leptonic} charge.  
We derive
analytic formulae for the computation of the baryon asymmetry that, in light of
these effects, are qualitatively different from those in the
established literature.  
In this scenario, for fixed CP-violating phases, the baryon asymmetry has opposite sign compared to that calculated using established formulae.

\end{abstract}
\maketitle
\section{Introduction}
\label{sec:intro}

Electroweak baryogenesis (EWB) is an attractive and testable explanation for the origin of the baryon asymmetry of the universe (BAU).
Characterized by the baryon-number-to-entropy-ratio $n_B/s$, the BAU
has been measured through studies of Big Bang Nucleosynthesis (BBN)
and the cosmic microwave background (CMB) to be in the following range
\be n_B/s = \left\{
\begin{array}{ccc}
(6.7 \; - \; 9.2) \times 10^{-11}  & \; & \textrm{BBN}  \\
(8.36  \; - \; 9.32) \times 10^{-11}  & \; & \textrm{CMB}  \end{array} \right.
\ee
at 95\% C.L.~\cite{Yao:2006px,Dunkley:2008ie}.
Assuming that the universe was matter-antimatter symmetric at some initial time (e.g. at the end of inflation), the creation of the BAU requires three conditions (the Sakharov conditions~\cite{Sakharov:1967dj}): (1) violation of baryon number, (2) violation of C and CP, and (3) either a departure from equilibrium or a violation of CPT.

In EWB, these conditions are realized in the following way.  First, a
departure from equilibrium is provided by a strongly first order
electroweak phase transition (EWPT) at temperature $T \sim 100$
GeV~\cite{Shaposhnikov:1987tw,Shaposhnikov:1987pf}.  During the EWPT,
bubbles of broken electroweak symmetry nucleate and expand in a
background of unbroken symmetry, filling the universe to complete the
phase transition.  Second, CP-violation may arise from complex phases.
These phases induce CP-violating interactions at the walls of the
expanding bubbles, where the Higgs vacuum expectation value (vev) is
time-dependent, leading the production of a CP-asymmetric charge
density.  This is the so-called CP-violating source.  This
CP-asymmetry, created for one species, diffuses ahead of the advancing
bubble and is converted into other species through inelastic
interactions in the plasma; in particular, some fraction is converted
into left-handed fermion charge density, denoted $n_L$.  Third, baryon
number is violated by non-perturbative $SU(2)_L$ processes, which are
unsuppressed outside the bubbles, in regions of unbroken electroweak
symmetry~\cite{'t Hooft:1976up,'t
Hooft:1976fv,Manton:1983nd,Klinkhamer:1984di,Kuzmin:1985mm}.
Following the common usage, we will refer to these as sphaleron
processes.  The presence of non-zero $n_L$ biases the sphaleron
processes, resulting in the production of a baryon
asymmetry~\cite{Cohen:1994ss}.  Electroweak sphalerons become quenched
once electroweak symmetry is broken, as long as the EWPT is strongly
first order; therefore, the baryon asymmetry becomes frozen in once it
is captured inside the expanding bubbles.

In this work, we consider the charge transport dynamics during the
EWPT: that is, how charge densities, induced by CP-violating sources,
diffuse, interact, and get converted into $n_L$, ultimately inducing
$n_B/s$.  Although in the Standard Model (SM) this dynamics are
insufficient to produce the observed BAU~\cite{nocpvinsm},
supersymmetric extensions of the SM can readily include all the
ingredients to make it successful\footnote{Another reason that EWB is
not viable in the SM is that there is no EWPT; for a Higgs mass $m_h
\ge 114$ GeV, electroweak symmetry is broken through a continuous
crossover~\cite{Kajantie:1996mn}.}.  The most commonly accepted supersymmetic scenario is the
following: the expanding bubble wall leads to a CP-violating source
for charge density in the Higgs sector, which is then converted into
third generation quarks through top Yukawa interactions, which in turn
is converted into quark charge density of all generations through
strong sphaleron processes~\cite{Huet:1995sh}.  The rate for baryon number production 
is proportional to $n_L$; in
this picture, $n_L$ receives contributions from left-handed quarks of
all three generations.

\begin{figure*}[t]
\includegraphics[scale=1]{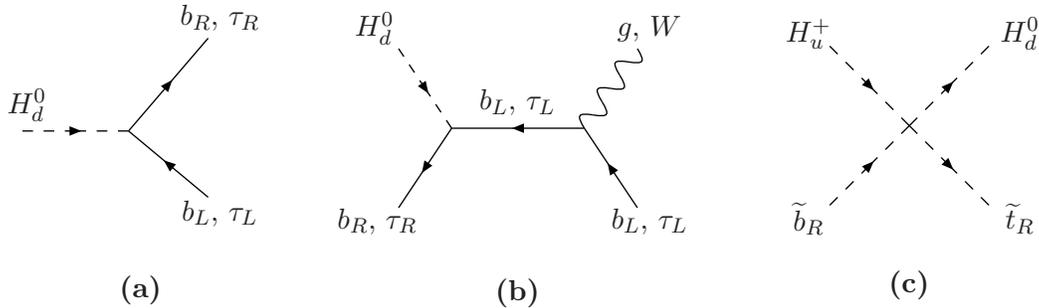}
\caption{Examples of bottom and tau Yukawa interactions from (a) absorption/decay, and (b) scattering processes involving an addition gauge boson, showing how Higgs density is converted into left-handed quark and lepton density.  Third generation Yukawa couplings also give rise to F-term-induced scattering processes (c).}
\label{fig:feyn}
\end{figure*}

However, as reported in a previous publication~\cite{Chung:2008ay}, we
have observed that bottom and tau Yukawa interactions, shown in Fig.~\ref{fig:feyn}, cannot in general be neglected from the computation of $n_{L}$
in supersymmetric EWB scenarios.  While bottom and tau Yukawa couplings are small
in the SM, in supersymmetric extensions they can be larger when the ratio of the vevs of the Higgs doublets, $v_u/v_d \equiv \tan\beta$,
is greater than unity. The inclusion of bottom and tau Yukawa interactions can change the EWB picture dramatically.
\begin{itemize}
\item {\it Quark charge density may be supressed:} For $\tan\beta \gtrsim 5$, bottom Yukawa interactions become
non-negligible, leading to two important effects: (1) strong sphaleron
processes no longer induce charge densities of first and second
generation quarks, and (2) the third generation left-handed quark
charge density vanishes when the masses of right-handed bottom and top
squarks are equal, or when their masses are large compared to the temperature $T$.
\item {\it Lepton charge density generated:} For $\tan\beta \gtrsim 20$, tau Yukawa interactions also are
non-negligible, leading to the conversion of Higgs charge density into
third generation lepton charge density\footnote{Measurements of the muon anomalous magnetic moment favors large $\tan\beta$; see, e.g.,  Ref.~\cite{Stockinger:2006zn}.}.  
\end{itemize}
These novel effects, which come into play for moderate $\tan\beta$, can lead to qualitatively different situations from those previously
considered for supersymmetric EWB.  In the present paper, we focus on a new
scenario, where $n_{L}$ can be purely leptonic.  We call this
``lepton-mediated electroweak baryogenesis.''  This scenario occurs when $\tan\beta \gtrsim 20$,
 and the right-handed bottom and top squark masses are, either, approximately equal ($m_{\widetilde b_R}\approx m_{\widetilde
t_R}$), or large compared to $T$ ($m_{\widetilde b_R}, \, m_{\widetilde
t_R} \gtrsim 500$ GeV).  (In other regions of parameter space where the quark contribution to $n_L$ is not quenched, the lepton contribution may still provide an additional enhancement or suppression to the total $n_L$.)  

In the lepton-mediated EWB scenario, the value of $n_B/s$ has opposite
sign compared to the value of $n_B/s$ computed when neglecting the
bottom and tau Yukawa rates.  The ingredients of this scenario will be
tested in the near future at the Large Hadron Collider and by
precision electric dipole moment (EDM)
searches~\cite{Pospelov:2005pr}.  Clearly, to the extent that these
experiments can determine the supersymmetric spectrum, and the signs and magnitudes of
relevant CP-violating phases, inclusion of these Yukawa rates may be essential for
testing the consistency of supersymmetric EWB with observation.

In Sec.~\ref{sec:simpboltzmann}, we present the system of Boltzmann
equations, generalized from previous work~\cite{Huet:1995sh,
Lee:2004we} to include bottom and tau Yukawa interactions.  In
Sec.~\ref{sec:analytic}, we provide an analytic estimate of the baryon
asymmetry in detail.  We solve the Boltzmann equations analytically in
the limit that $\tan\beta \gg 1$, such that bottom and tau Yukawa
interactions are in chemical equilibrium.  A new qualitative feature
of our analysis is our treatment of lepton diffusion; we argue
analytically how the left-handed lepton charge density (and therefore
$n_B/s$) is enhanced by virtue of right-handed leptons diffusing more
efficiently in the plasma.

In Sec.~\ref{sec:numerical}, we verify these conclusions numerically.
First, we calculate the bottom and tau Yukawa interaction rates,
showing in what regimes they are sufficiently fast to induce chemical
equilibrium.  Next, after defining the parameters of our
lepton-mediated EWB scenario, we solve the system of Boltzmann
equations numerically.  We illustrate all of the aforementioned new
effects and verify the agreement between our numerical and analytical
solutions.  In Sec.~\ref{sec:conclude}, we summarize our results.  The
Appendix summarizes some additional numerical inputs used for this
work.

\section{Boltzmann Equations} 
\label{sec:simpboltzmann}

\subsection{Preliminaries}

The transport dynamics leading to CP-asymmetric charge densities during the EWPT are governed by a system of Boltzmann equations.
These Boltzman equations have been derived using the closed-time-path formulation of non-equilibrium quantum field
theory~\cite{Chou:1984es}, leading to a system of equations of the
form \be
\label{eq:trans1}
\partial_\mu\ j_i^\mu = - \frac{T^2}{6} \sum_X \Gamma_{X} \left( \,
\mu_i + \mu_j + ... - \mu_k - \mu_\ell - ... \frac{}{} \right) +
S_i^\cpv \ee 
where $j^\mu_i$ is the charge current density of the
species $i$.  The density $j^\mu_i$, induced by CP-violating source $S^\cpv_i$, is coupled to other species via coefficients
$\Gamma_X$ that describe the rate for a process $i + j +
... \leftrightarrow k + \ell + ...$ to occur.  (We have explicitly
factored $T^2/6$ out of $\Gamma_X$, for reasons that will become clear
below.)  The chemical potentials are denoted by $\mu_i$. Chemical
equilibrium, occuring when
\be
\label{eq:chemeq}
\mu_i + \mu_j + ... - \mu_k - \mu_\ell - ... = 0  \; ,
\ee
is maintained when when the interaction rate $\Gamma_X$ is sufficiently large.

Following previous
work~\cite{Huet:1995sh,Cline:2000nw,Lee:2004we,Cirigliano:2006wh}, we
simplify Eq.~\eqref{eq:trans1} in three ways.  First, we assume a
planar bubble wall profile, so that all charge densities are
functions only of $z$, the displacement from the moving bubble wall in
its rest frame.  Second, we apply Fick's
law~\cite{Fick:1855a,Fick:1855b,Joyce:1994bi,Cohen:1994ss,Joyce:1994zn},
which allows us to replace $\mathbf{j}_{\,i} \to - D_i \nabla n_i$ on the
LHS of Eq.~\eqref{eq:trans1}, with charge density $n_i \equiv
j_i^{\, 0}$.  The diffusion constant $D_i$ is the mean free path of particle $i$ in the plasma.  Third, the chemical
potentials appearing in Eq.~\eqref{eq:trans1} are related to their
corresponding charge densities by \be n_i = \frac{T^2}{6} \, k_i \,
\mu_i + \mathcal{O}\left(\frac{\mu_i}{T}\right)^3\;, \label{eq:nmu}
\ee where we have performed an expansion assuming $\mu_i/T \ll 1$.  In the above, the statistical weight
$k_i$ is defined by 
\be k_i = g_i \, \frac{6}{\pi^2}\,
\int_{m_i/T}^\infty\, dx \, x\ \frac{e^x}{(e^x\pm1)^2}\,
\sqrt{x^2-m_i^2/T^2}\ \ \, , \ee 
in which $g_i$ counts the number of
internal degrees of freedom, the $+$ ($-$) sign is taken for fermions (bosons),
and the mass of the $i$th particle $m_i$ is taken to be the effective
mass at temperature $T$.  In our analysis to follow, these $k$-factors are ubiquitous; they essentially count
the degrees of freedom of a species in the plasma, weighted by a Boltzmann suppression.  

Through these three
simplifications, the Boltzmann equations become a system of
coupled, second order, ordinary differential equations for the set of
charge densities $n_i(z)$.  Ultimately, it is the total left-handed
fermionic charge density 
\be n_{L} \equiv \sum_{i=1}^3 \left( n_{u_L^i}
+ n_{d_L^i} + n_{\nu_L^i} + n_{\ell_L^i} \right) \ee 
that biases weak
sphaleron transitions, thereby determining $n_B/s$.

While in principle there is an interaction coefficient $\Gamma_X$ for
every interaction in the MSSM Lagrangian, we can determine which ones
need to be taken into account for the computation of $n_{L}$ by
considering the relevant time scales. After a time $t$, charge
densities created at the bubble wall will have diffused on average a distance
$d_{\textrm{diff}} = \sqrt{\bar{D} \, t}$ (with the effective
diffusion constant $\bar{D}$ to be defined below).  At the same time,
the moving bubble wall advances a distance $d_{\textrm{wall}} = v_w \,
t$.  The diffusion time scale, defined by $d_{\textrm{diff}} =
d_{\textrm{wall}}$, gives the time that it takes for charge, having
been created at the bubble wall and having diffused into the unbroken
phase, to be recaptured by the advancing bubble wall and be quenched
through CP-conserving scattering within the phase of broken
electroweak symmetry.  This time scale is 
\be 
\tau_{\rm diff} \equiv \bar{D}/v_w^2 \;.
\ee 
Numerically, we have $\tau_{\rm diff} \sim 10^4/T$ (shown in Sec.~\ref{sec:numerical}).
To this, we compare $\tau_X \equiv \Gamma_X^{-1}$, the interaction time scale. 
If $\tau_X \gg \tau_{\rm diff}$, then the process $i + j +
... \leftrightarrow k + \ell + ...$ is slow and $\Gamma_X$ may be
neglected from the Boltzmann equations.  Physically speaking, charge
density is recaptured by the advancing bubble wall before conversion
processes can occur.  On the other hand, if $\tau_X \ll \tau_{\rm
diff}$, then these interactions are rapidly occuring as the charge
density is diffusing ahead of the advancing wall, leading to chemical
equilibrium \eqref{eq:chemeq}.  Expressed in terms of charge densities, the 
chemical equilibrium condition is
\be
\frac{n_i}{k_i} + \frac{n_j}{k_j} + \; ... \; - \frac{n_k}{k_k} - \frac{n_\ell}{k_\ell} \; - ... \; = 0 \; . \label{eq:chemeq2}
\ee
In this case, the interaction
$\Gamma_X$ must be included in the Boltzmann equations.

A similar argument tells us how we expect deviations from Eq.~\eqref{eq:chemeq2} to arise.  Suppose that species $i$ is produced from the expanding bubble wall at $z=0$.  On distance scales $|z| \lesssim \sqrt{\bar{D} \, \tau_X}$, close to the bubble wall, Eq.~\eqref{eq:chemeq2} will break down: particles $i$ have not had enough time to interact via $\Gamma_X$.

\subsection{Setting up the Boltzmann equations}

We now derive the Boltzmann equations within the context of the Minimal Supersymmetric Standard Model (MSSM).
In principle, the complete system of Boltzmann equations encompasses
one equation for each species of particle.  
However, the assumption
that certain interactions $\Gamma_X$ are in chemical equilibrium (such
that $\tau_{\rm diff} \gg \tau_X$) implies relations among the
relevant chemical potentials (and therefore among their corresponding
charge densities), allowing one to reduce the system.  First, we assume that 
weak interactions (neglecting flavor mixing) are in chemical equilibrium, so that particles in the same isodoublet have equal chemical potential.
Second, we assume that gaugino interactions (involving SM particles and their superpartners) are also in chemical equilibrium, so that a particle and its superpartner have equal chemical potential~\cite{superequilibrium}.

Under these assumptions, the complete set of charge
densities relevant for the computation of $n_B/s$ is 
\begin{align} 
U_i &\equiv n_{u^i_R} + n_{\widetilde u^i_R}\; ,  & Q_i &\equiv n_{u^i_L} + n_{d^i_L} + n_{\widetilde u^i_L} + n_{\widetilde d^i_L}  \;, \notag \\ 
D_i &\equiv n_{d^i_R} + n_{\widetilde d^i_R} \label{eq:densities2} \; ,  & H &\equiv n_{H_u^+} + n_{H_u^0} - n_{H_d^-} - n_{H_d^0} + n_{\widetilde H^\pm} + n_{\widetilde H^0} \; , \\
R_i &\equiv n_{e^i_R} + n_{\widetilde e^i_R} \; , &
L_i &\equiv n_{\nu^i_L} + n_{e^i_L} + n_{\widetilde \nu^i_L} + n_{\widetilde e^i_L} \; ,  \notag 
\end{align}
where $i \in \{1,2,3\}\, $ labels the generations.  Furthermore, we define
the following additional notation: $Q \equiv Q_3$, $T \equiv U_3$, $B
\equiv D_3$, $L \equiv L_3$, and $R \equiv R_3$.

The system of Boltzmann equations contains, in principle, a
coefficient $\Gamma_X$ for every interaction in the MSSM.  However,
interactions that satisfy $\tau_X \gg \tau_{\rm diff}$ may be
neglected.  In particular, we neglect interactions induced by first
and second generation quark and lepton Yukawa couplings.  The weak
sphaleron rate $\Gamma_{ws}$ may also be neglected, since $\tau_{ws} \sim 10^5/T
\gg \tau_{\rm diff}$ \cite{Bodeker:1999gx}.  Therefore, baryon and lepton number are
conserved in the collision terms of the Boltzmann equations.  

Not all of the densities in Eq.~\eqref{eq:densities2} are
independent.    
Neglecting electroweak sphalerons from the Boltzmann equations, baryon and lepton number are individually conserved:
\be
\int^\infty_{-\infty} dz \; \sum_{i=1}^3 \, \left( Q_i + U_i + D_i \right) = \int^\infty_{-\infty} dz \; \sum_{i=1}^3 \, \left(  L_i + R_i \right) = 0 \;.
\ee
Because the left- and right-handed (s)lepton have different gauge quantum numbers, they have different diffusion constants in the plasma.  Even though lepton number is {\it globally conserved}, regions of net lepton number can develop since $R$ diffuses more easily than $L$ since right-handed (s)leptons do not undergo SU(2) gauge interactions. 
For quarks and squarks, this does not occur since the left- and right-handed (s)quark diffusion constants, dominated by strong interactions, are approximately equal~\cite{Joyce:1994zn}.  Therefore, baryon number is {\it locally} conserved:
\be 
\sum_{i=1}^3 \, \left( Q_i + U_i + D_i \right) = 0 \;. \label{eq:blcons} 
\ee 

Other simplifications arise since we neglect first and second generation Yukawa couplings.
There is no production of first and second generation
lepton charge, so $L_{1,2} = R_{1,2} = 0$.  Next,
first and second
generation quark charge can only be produced through strong
sphaleron processes, e.g., $t_L \, \bar t_R \overset{\textrm{ss}}\longrightarrow \bar b_L \, b_R  \sum_{i=1,2} \bar u^i_L \, u^i_R \, \bar d^i_L \, d^i_R$, changing the number of left- and right-handed quarks by one unit per flavor.
Since first and second generation quarks are produced in equal numbers, we have 
\be Q_1 = Q_2 = - 2 \, U_1 = -2 \, U_2 =
- 2 \, D_1 = -2 \, D_2 \;. \label{eq:ss} \ee 
Together, Eqs.~(\ref{eq:blcons},\ref{eq:ss}) imply
that 
\be B = - (T+Q) \;. \label{eq:btq} \ee 
Therefore, we may consider
a reduced set of Boltzmann equations involving only the densities $Q$,
 $T$, $Q_1$, $L$ , $R$, $H$; the remaining densities are then
determined by Eqs.~(\ref{eq:ss},\ref{eq:btq}). The equations are
of the form of Eq.~(\ref{eq:trans1}), where we use the
relation given in Eq.~(\ref{eq:nmu}) to express the chemical potentials in terms of
charge densities.  For the quarks and squarks, we obtain
\begin{widetext}
\begin{subequations}
\label{eq:newbe2}
\bea
v_w \, Q^\prime - D_Q \, Q^{\prime\prime} &=& \; - \; \Gamma_{yt} \left( \frac{Q}{k_Q} - \frac{T}{k_T} + \frac{H}{k_H} \right) - \; \Gamma_{yb} \left( \frac{Q}{k_Q} + \frac{T+Q}{k_B} - \frac{H}{k_H} \right) \label{eq:Qeqn}\\
&\;& \; - \;\Gamma_{mt} \left(\frac{Q}{k_Q} - \frac{T}{k_T} \right) -  \Gamma_{mb} \left(\frac{Q}{k_Q} + \frac{T+Q}{k_B} \right) - S_{\widetilde t}^\cpv - S_{\widetilde b}^\cpv \notag\\
&\;& \;  - 2 \, \Gamma_{ss} \left( 2 \, \frac{Q}{k_Q} - \frac{T}{k_T} + \frac{Q+T}{k_B} + \frac{1}{2} \sum_{i=1}^2 \left[ 4 \, \frac{1}{k_{Q_i}} + \frac{1}{k_{U_i}} + \frac{1}{k_{D_i}} \right] \, Q_1 \; \right) \notag \\
v_w \, T^\prime - D_Q \, T^{\prime\prime} &=& \; \Gamma_{yt} \left( \frac{Q}{k_Q} - \frac{T}{k_T} + \frac{H}{k_H} \right) + \Gamma_{mt} \left(\frac{Q}{k_Q} - \frac{T}{k_T} \right) + S_{\widetilde t}^\cpv \label{eq:Teqn}\\
&\;& \;  + \; \Gamma_{ss} \left( 2 \, \frac{Q}{k_Q} - \frac{T}{k_T} + \frac{Q+T}{k_B} + \frac{1}{2} \sum_{i=1}^2 \left[ 4 \, \frac{1}{k_{Q_i}} + \frac{1}{k_{U_i}} + \frac{1}{k_{D_i}} \right] \, Q_1 \; \right) \notag \\
\label{eq:Q1eqn}
v_w \, Q_{1}^\prime - D_Q \, Q_{1}^{\prime\prime} &=& \;  - 2 \, \Gamma_{ss} \left( 2 \, \frac{Q}{k_Q} - \frac{T}{k_T} + \frac{T+Q}{k_B} + \frac{1}{2} \sum_{i=1}^2 \left[ 4 \, \frac{1}{k_{Q_i}} + \frac{1}{k_{U_i}} + \frac{1}{k_{D_i}} \right] \, Q_1 \; \right) \; ;
\eea
\end{subequations}
and for Higgs bosons and Higgsinos we have
\bea
\label{eq:higgs}
v_w \, H^\prime - D_H \, H^{\prime\prime} &=& \; - \; \Gamma_{yt} \left( \frac{Q}{k_Q} - \frac{T}{k_T} + \frac{H}{k_H} \right) - \Gamma_h \, \frac{H}{k_H} + S_{\widetilde H}^\cpv \label{eq:Heqn} \\
&\;& \; + \; \Gamma_{yb} \left( \frac{Q}{k_Q} + \frac{Q+T}{k_B} - \frac{H}{k_H} \right) + \Gamma_{y\tau} \left( \frac{L}{k_L} - \frac{R}{k_R} - \frac{H}{k_H} \right) \notag\; ;
\eea
and lastly for leptons and sleptons we have
\begin{subequations}
\label{eq:lepeqn}
\bea
v_w \, L^\prime - D_L \, L^{\prime\prime} &=& \; - \; \Gamma_{y\tau} \left( \frac{L}{k_L} - \frac{R}{k_R} - \frac{H}{k_H} \right) - \Gamma_{m\tau} \left(\frac{L}{k_L} - \frac{R}{k_R} \right) - S_{\widetilde \tau}^\cpv \label{eq:Leqn}\\
v_w \, R^\prime - D_R \, R^{\prime\prime} &=& \; \Gamma_{y\tau} \left( \frac{L}{k_L} - \frac{R}{k_R} - \frac{H}{k_H} \right) + \Gamma_{m\tau} \left(\frac{L}{k_L} - \frac{R}{k_R} \right) + S_{\widetilde \tau}^\cpv \label{eq:Reqn} \;.
\eea
\end{subequations}
\end{widetext}
The relevant interaction coefficients in Eqs.~(\ref{eq:newbe2}-\ref{eq:lepeqn}) are as follows:
\begin{itemize}
\item The coefficients $\Gamma_{yi}$, where $i \in \{t,b,\tau\}$, denote the interaction rates arising from third generation Yukawa couplings $y_i$.  (The top Yukawa interaction rate has been denoted $\Gamma_y$ in previous work.)
\item The strong sphaleron rate is $\Gamma_\mathrm{ss}=16 \, \kappa^\prime \, \alpha_s^4 \, T$, where $\alpha_s$
is the strong coupling and $\kappa^\prime\sim\mathcal{O}(1)$ \cite{Moore:1997im}.
\item The coefficients $\Gamma_h$ and $\Gamma_{mi}$, where $i \in \{t,b,\tau\}$, denote the CP-conserving scattering rates of particles with the background Higgs field within the bubble~\cite{Lee:2004we}.
\end{itemize}
We also allow for new CP-violating sources $S_{\widetilde b, \widetilde \tau}^\cpv$, although in the present work we do not evaluate their magnitudes.  In the MSSM, the most viable CP-violating source is $S_{\widetilde H}^\cpv$, arising from CP-violating Higgsino-Bino mixing within the expanding bubble wall~\cite{Li:2008ez}; in our work, we take this as the sole source of CP-violation.
The constant $v_w \simeq 0.05$ is the velocity of the expanding bubble wall~\cite{bubbles}.  The $k$-factors, e.g.
\be
\label{eq:ksusy}
k_R \equiv  k_{\tau_R} + k_{\widetilde\tau_R} \; , \quad 
k_Q \equiv k_{t_L} + k_{b_L} + k_{\widetilde t_L} + k_{\widetilde b_L}, \; \ldots \; ,
\ee
follow the same notation as in Eqs.~\eqref{eq:densities2}.

After solving the system of Boltzmann equations
(\ref{eq:newbe2}-\ref{eq:lepeqn}) for each density, the left-handed fermion
charge density is \be
\label{eq:nL}
n_{L} = \left(\frac{k_q}{k_Q}\right) \, Q + \sum_{i=1,2}
\left(\frac{k_{q_i}}{k_{Q_i}} \right) \, Q_1 +
\left(\frac{k_\ell}{k_L}\right) \, L \;, \ee 
where $k_q \equiv k_{t_L}
+ k_{b_L}$, $k_\ell \equiv k_{\nu^\tau_L} + k_{\tau_L}$, {\it etc.}
The three terms in Eq.~\eqref{eq:nL} correspond to the contributions
to $n_{L}$ from third generation quarks, first/second generation quarks,
and third generation leptons, respectively.  If the masses of all
left-handed squarks and sleptons are much above the temperature of the
phase transition, only fermions contribute to the left-handed density
and we have \be n_{L} \simeq Q + 2 \, Q_1 + L \;. \label{eq:nL2} \ee

Finally, we show how our Boltzmann equations reproduce those given
in previous work in the limit $y_{b}, \, y_\tau \to 0$.  In this
limit, we can neglect the rates $\Gamma_{yb,\tau}$ and
$\Gamma_{mb,\tau}$, and CP-violating sources $S_{\widetilde b,
\widetilde \tau}^\cpv$.  First, since there is no source for lepton
charge, we have $L = R = 0$.  Second, the only source for $B$ density
is strong sphaleron processes; therefore, we have \be \label{eq:b2q1}
-2\, B = Q_1 \ee in analogy with Eq.~\eqref{eq:ss}.  Thus,
Eqs.~(\ref{eq:btq},\ref{eq:b2q1}) imply that $Q_1 = 2(Q+T)$.
Therefore, by Eq.~\eqref{eq:nL2}, we have the often-used relation
$n_{L} = 5 Q + 4 T$; this relation is no longer valid for
$\tau_{yb},\tau_{y\tau} \lesssim \tau_{\rm diff}$.  In addition, the
Boltzmann equations of
Refs.~\cite{Huet:1995sh,Lee:2004we,Cirigliano:2006wh} follow from
Eqs.~(\ref{eq:Qeqn},b,\ref{eq:Heqn}); they too are no longer valid
for $\tau_{yb},\tau_{y\tau} \lesssim \tau_{\rm diff}$.


\section{Analytic results}
\label{sec:analytic}

In this section, we estimate the solution to the Boltzmann equations,
Eqs.~(\ref{eq:newbe2}-\ref{eq:lepeqn}), of which the endpoint is an
expression for $n_L$, the left-handed fermion charge density that
biases weak sphalerons.  We assume that top, bottom, and tau Yukawa
interactions, in addition to strong sphaleron and gaugino
interactions, are all in chemical equilibrium. These assumptions lead
to a series of conditions relating the chemical potentials, and
therefore the number densities, of various species. By exploiting
these relations, we will express all quark and lepton densities $Q, \,
T, \, Q_1, \, L, \, R$ in terms of the Higgs density $H$; then, we
will simplify the full system of Boltzmann equations to a single
equation for $H$, which is analytically solvable~\cite{Huet:1995sh}.

\subsection{Lepton charge densities}

When tau Yukawa interactions are in chemical equilibrium condition, the relation
\be
\frac{L}{k_L} - \frac{H}{k_H} - \frac{R}{k_R} =0 \;. \label{eq:taueq}
\ee
is satisfied.
The sum of the Boltzmann equations for $L$ and $R$ \eqref{eq:lepeqn} is
\be
v_w \, (R+L)^\prime - \left( D_R \, R^{\prime\prime} + D_L \, L^{\prime\prime} \right) = 0 \;. \label{eq:R+L}
\ee
Since the left- and right-handed lepton diffusion constants are not equal, there is no simple relation that would allow us to relate $R$ to $L$.  However, in the static limit (where $v_w \to 0$), Eq.~\eqref{eq:R+L} implies that
\be
D_L \, L = - \, D_R \, R \;.
\ee
(We have assumed the boundary conditions $L(\infty) = L^\prime(\infty) = R(\infty) = R^\prime(\infty) = 0$.)  Therefore, we have
\begin{subequations}
\label{eq:LRH}
\bea
L(z) &\equiv& \kappa_L \, H(z) + \Delta L(z)= \frac{k_L}{k_H} \, \frac{D_R \, k_R}{D_L \, k_L + D_R \, k_R} \, H(z)+ \Delta L(z) \label{eq:LRHa}\\ 
R(z) &\equiv& \kappa_R \, H(z) + \Delta R(z) = - \frac{k_R}{k_H} \, \frac{D_L \, k_L}{D_L \, k_L + D_R \, k_R} \, H(z) + \Delta R(z) \;,
\eea
\end{subequations}
where $\Delta L$ and $\Delta R$ are the corrections to these
relations, derived below.

Let us now describe the physics of Eqs.~\eqref{eq:LRH} through two limiting cases.  Case (i): set $D_R = D_L$. In this limit, Eq.~\eqref{eq:R+L} implies that lepton number is {\it locally} conserved: $L+R = 0$.  The Higgs density $H$, created by the CP-violating source, is converted into $L$ through tau Yukawa interactions, until chemical equilibrium \eqref{eq:taueq} is reached, when
\be
L(z) = \frac{ k_L}{k_H}  \, \frac{k_R }{k_L + k_R } \, H(z) \;. \label{eq:LwDReDL}
\ee
Case (ii): take $D_R \to \infty$, keeping $D_L$ finite.  Any $R$ density created by tau Yukawa interactions instantly diffuses away to $z = \pm \infty$; therefore, we set $R=0$.  Now, tau Yukawa chemical equilibrium \eqref{eq:taueq} implies
\be
L(z) = \frac{k_L}{k_H} \, H(z) \;. \label{eq:LwDRinf}
\ee
In other words, tau Yukawa interactions will enforce chemical equilibrium locally.  Since the RH lepton density is diffusing away, reducing the local $R$, more conversion of $H$ into $R$ and $L$ occurs to compensate, thereby resulting in more LH lepton density.  This conversion ceases when Eq.~\eqref{eq:LwDRinf} is reached.  Therefore, a large diffusion constant for RH leptons enhances the density for LH leptons.  This enhancement, maximized for $D_R \to \infty$, is at most a factor of
\be
\frac{k_R + k_L}{k_R} \sim \mathcal{O}(5) \; .
\ee
Both cases agree with Eqs.~\eqref{eq:LRH}, setting $\Delta L, \, \Delta R \to 0$.

Next, consider the case of physical relevance, where $D_R \gg D_L$,
but keeping both $D_R, D_L$ finite.  Close to the bubble wall, LH
lepton density will be enhanced, as argued above.  However, far from
the bubble wall, an additional effect occurs: RH lepton density,
having diffused far into the unbroken phase, is converted into $L$ and
$H$ by tau Yukawa interactions.  This effect {\it suppresses} $L$.
Close to the bubble wall, Higgsinos created by the CP-violating source
($H>0$) will be converted into LH leptons ($L>0$) and RH anti-leptons
($R<0$), and then, far from the wall, the RH anti-leptons will be
converted into LH anti-leptons, thereby suppressing $L$.  This physics
is incorporated in the non-local corrections $\Delta L$ and $\Delta
R$, which we now consider.  Using Eqs.~(\ref{eq:taueq}, \ref{eq:R+L},
\ref{eq:LRH}), we can derive differential equations for these
densities:
\begin{subequations}
\bea
- \, D_{LR} \, \Delta L^{\prime\prime} + v_w \, \Delta L^\prime &=& v_w \, \frac{k_R \, k_L^2}{k_H (k_L+k_R)^2} \, \frac{D_L - D_R}{D_{LR}} \, H^\prime \\
- \, D_{LR} \, \Delta R^{\prime\prime} + v_w \, \Delta R^\prime &=& v_w \, \frac{k^2_R \, k_L}{k_H (k_L+k_R)^2} \, \frac{D_L - D_R}{D_{LR}} \, H^\prime \;,
\eea
\end{subequations}
where $D_{LR} \equiv (D_L \, k_L + D_R \, k_R)/(k_L+k_R)$.
With the boundary conditions $\Delta L(\pm \infty) = \Delta R(\pm \infty) = 0$, the solutions to these equations are
\begin{subequations}
\bea
\Delta L(z) &=& v_w \, \frac{k_L^2 \, k_R}{k_H \, (k_R + k_L)^2 } \, \frac{D_L - D_R}{D_{LR}^2} \, \int_{z}^\infty dz^\prime \: H(z^\prime) \, e^{v_w(z-z^\prime)/D_{LR} } \\
\Delta R(z) &=& v_w \, \frac{k_L \, k_R^2}{k_H \, (k_R + k_L)^2 } \, \frac{D_L - D_R}{D_{LR}^2} \, \int_{z}^\infty dz^\prime \: H(z^\prime) \, e^{v_w(z-z^\prime)/D_{LR} } \; .
\eea
\label{eq:DeltaLR}
\end{subequations}
These terms describe how regions of net lepton number can develop when $D_R \ne D_L$.  Using Eqs.~(\ref{eq:LRH},\ref{eq:DeltaLR}), it is straight-forward to show that
these solutions for $L$ and $R$ satisfy
\be
\int^\infty_{-\infty} dz \, (L + R) = 0 \; ,
\ee
conserving lepton number.

In our numerical study, we find that the impact from $\Delta L$ and $\Delta R$ on the analytic
computation of $n_B/s$ (given below) is only $\mathcal{O}(10\%)$.  Since there are
much larger uncertainties in the analytic computation, it is safe to
neglect the non-local terms $\Delta L$ and $\Delta R$ from
Eqs.~\eqref{eq:LRH}.

\subsection{Quark charge densities}

When top and bottom Yukawa interactions are in chemical equilibrium, the relations
\begin{subequations}\label{eq:relations}
\bea
\frac{Q}{k_Q} + \frac{H}{k_H} - \frac{T}{k_T} = 0\,, \\
\frac{Q}{k_Q} - \frac{H}{k_H} + \frac{B}{k_B} = 0\, .
\eea
\end{subequations}
are satisfied; {\it cf.} Eqns.~(\ref{eq:chemeq},~\ref{eq:nmu}).
These equations imply that
\be
2 \, \frac{Q}{k_Q} - \frac{T}{k_T} - \frac{B}{k_B} = 0 \;. \label{eq:vanish}
\ee
First and second generation quark densities only couple to third generation densities, via strong sphaleron interactions, through the linear combination $(2Q/k_Q - T/k_T - B/k_B)$, as can be seen from Eqns.~(\ref{eq:newbe2}).  Since this combination vanishes, third generation quark densities do not induce 1st/2nd generation quark densities.  Mathematically, if we impose Eq.~\eqref{eq:vanish}, the $Q_1$ Boltzmann equation \eqref{eq:Q1eqn} becomes
\be
v_w \ Q_1^\prime - D_q \, Q_1^{\prime\prime} \; \propto \; - \; \Gamma_{ss}  \, Q_1 \;,
\ee
which, with the boundary conditions $Q_1(\pm\infty) = 0$, implies $Q_1(z) = 0$.  According to Eq.~\eqref{eq:blcons}, we have $U_i = D_i = - Q_i/2 = 0$, for $i=1,2$.  
Therefore, we conclude that all first and second generation quark and squark charge densities vanish in the presence of fast top and bottom Yukawa interations.  Strong sphalerons only induce first and second generation densities in order to wash out an asymmetry between left- and right-handed quark chemical potentials; when bottom Yukawas are active, this asymmetry vanishes and strong sphalerons have no effect.

Eqns.~(\ref{eq:btq},\ref{eq:relations}) imply
\bea
T &\equiv& \kappa_T \, H = \frac{k_T}{k_H} \, \frac{2 k_B + k_Q}{k_B+k_Q+k_T} \, H\notag\\
Q &\equiv& \kappa_Q \, H = \frac{k_Q}{k_H} \, \frac{k_B - k_T}{k_B+k_Q+k_T} \, H \label{eq:kappas} \\
B &\equiv& \kappa_B \, H = - \, \frac{k_B}{k_H} \, \frac{2 k_T + k_Q}{k_B+k_Q+k_T} \, H \notag \;.
\eea

The contribution to $n_L$ from third generation LH quarks is
\be
n_{u_L^3} + n_{d_L^3} = \frac{k_q}{k_H} \, \frac{k_B - k_T}{k_B+k_Q+k_T} \, H \;,
\ee
while that from first and second generation LH quarks vanishes.  Let us contrast these results to previous work that neglected bottom Yukawa interactions~\cite{Huet:1995sh}:
\begin{subequations}
\bea
n_{u_L^3} + n_{d_L^3} &=& \frac{k_q}{k_H} \, \frac{k_B - 9 k_T}{k_B+9 k_Q+9 k_T} \: H \;, \\
n_{u_L^{i}} + n_{d_L^{i}} &=& \frac{k_{q_{i}}}{k_H} \, \frac{2k_Q(k_B - 9 k_T) + 2 k_T(9k_T + 2k_B)}{k_B+9 k_Q+9 k_T} \: H \;, \quad i=1,2 \; ,
\eea
\end{subequations}
The formulae are completely different.  Whereas in previous work significant baryon asymmetry could arise from first and second generation LH quarks, the presence of bottom Yukawa interactions completely changes the picture: no first and second generation quark density is created.  
In addition, with fast bottom Yukawa interactions, the third generation quark charge vanishes when $k_T \simeq k_B$, or equivalently $m_{\widetilde t_R} \simeq m_{\widetilde b_R}$; without them, this cancellation never occurs.  

Let us explain the physical origin of this cancellation.  
Suppose that the CP-violating source creates positive Higgs/Higgsino density, such that $H>0$.  Due to hypercharge conservation, top Yukawa interactions convert Higgsinos and Higgs bosons into LH quark and squark antiparticles (driving $Q<0$) and RH top quark and squark particles (driving $T>0$), while bottom Yukawa interactions drive $Q>0$ and $B<0$.  
Which effect wins is determined by whether $T$ or $B$ has more degrees of freedom available, according to the equipartition theorem.  This is determined by the statistical weights $k_T$ and $k_B$, which are governed by the masses $m_{\widetilde t_R}$ and $m_{\widetilde b_R}$.  When the masses are equal, we have $k_B \simeq k_T$, suppressing $n_{u_L^3}+n_{d_L^3}$.  Similarly, the sign of $(n_{u_L^3}+n_{d_L^3})/H$ is positive or negative, depending on whether $m_{\widetilde t_R}/m_{\widetilde b_R}$ is greater or less than unity, respectively.
In the lepton-mediated scenario, we suppress the quark contribution by choosing $m_{\widetilde t_R} \simeq m_{\widetilde b_R}$.
In scenarios beyond the MSSM, it is also suppressed for $m_{\widetilde t_R}, \, m_{\widetilde b_R} \gg T$.

\subsection{Solving the Boltzmann equation}

In terms of $H$, the left-handed fermion charge density \eqref{eq:nL} becomes
\be
\label{eq:lhcharge}
n_L(z) = \frac{k_q}{k_H} \, \frac{k_B - k_T}{k_B + k_Q + k_T} \: H(z) + \frac{k_\ell}{k_H} \, \frac{k_R \, D_R}{k_L \, D_L  + k_R \, D_R} \: H(z)  + \frac{k_\ell}{k_L} \; \Delta L(z) \;,
\ee
where $\Delta L$ is given in Eq.~\eqref{eq:LRHa}.  
The first term is the contribution to $n_L$ from third generation quarks, while the second and third terms are contributions from third generation leptons.  The lepton contribution is predominantly given by the second term only; the third term, as discussed above, is suppressed for $v_w \ll 1$.
This equation is the main result of this paper; from it, we infer several conclusions:
\begin{itemize}
\item The lepton contribution is enhanced for $m_{\widetilde\tau_R}
\ll m_{\widetilde\tau_L}$, when $k_R$ is largest and $k_L$ smallest;
~({\it cf.} Eqs.~(\ref{eq:nmu},\ref{eq:ksusy})).  It is also enhanced
for $D_R \gg D_L$.  Its sign is fixed with respect to $H$, which in
turn is fixed by the sign of the CP-violating source, as we show
below.  Therefore, in a lepton-mediated EWB scenario, where $n_L$ is
predominantly leptonic, the sign of the CP-violating phase most
relevant for EWB uniquely fixes the sign of $n_B/s$, in contrast with
the quark-mediated scenarios.
\item Left-handed charge arises from third generation quarks and leptons, and not first and second generation quarks and leptons.  The form of $n_L$ is qualitatively different than in previous treatments that neglected $\Gamma_{yb}$ and $\Gamma_{y\tau}$, where left-handed charge came from quarks of all generations, and not from leptons.  
\item Furthermore, the quark contribution to $n_L$ vanishes for $m_{\widetilde t_R} = m_{\widetilde b_R}$, since $k_B = k_T$.  Its sign is opposite to that of the leptonic contribution for $m_{\widetilde t_R} < m_{\widetilde b_R}$ and the same for $m_{\widetilde t_R} > m_{\widetilde b_R}$.
\end{itemize}
We explore these implications in more detail numerically in Sec.~\ref{sec:numerical}.  

We emphasize that our conclusions are quite general, although it appears that our Boltzmann equations (\ref{eq:newbe2}-\ref{eq:lepeqn}) have been specialized to the MSSM.  In any extention of the MSSM, Eq.~\eqref{eq:lhcharge} and its conclusions remain valid if the following conditions hold: (i) third generation Yukawa interaction rates are faster than the diffusion rate, and (ii) CP-violation is communicated to the first and second generation quark sectors solely through strong sphalerons.

Since $n_B/s$ is determined by $n_L$, all that remains is to solve for the Higgs charge density $H$.  We can reduce the Boltzmann equations (\ref{eq:newbe2}-\ref{eq:lepeqn}) into a single equation for $H$ by taking the appropriate linear combination of equations
\be
\eqref{eq:Qeqn} + 2 \times \eqref{eq:Teqn} +  \eqref{eq:Heqn} + \eqref{eq:Leqn} \;,
\ee
such that the Yukawa and strong sphaleron rates all cancel, and expressing the densities $L,Q,T$ in terms of $H$ using Eqs.~(\ref{eq:LRH},\ref{eq:kappas}).  This master Boltzmann equation equation is an integro-differential equation for $H(z)$, due to the presence of the $\Delta L$ term.  Therefore, for simplicity, we treat $\Delta L$ perturbatively: first, we neglect $\Delta L$ in our solution for $H$, and then, given our solution $H$, we include the $\Delta L$ contribution in Eq.~\eqref{eq:lhcharge} for $n_L$.  Neglecting $\Delta L$, the master Boltzmann equation is
\be
\label{eq:Hdiff}
v_w \, H^\prime - \bar{D} \, H^{\prime\prime} = \; - \; \bar{\Gamma} \, H + \bar{S} \;,
\ee
where
\begin{subequations}
\label{eq:bar}
\be
\bar{D} = \frac{D_H + D_Q (\kappa_T - \kappa_B) + D_L \, \kappa_L }{1 + \kappa_T - \kappa_B + \kappa_L} 
\ee
\be
\bar{\Gamma} = \frac{ \Gamma_h + \Gamma_{mt} + \Gamma_{mb}  + \Gamma_{m\tau} }{k_H(1 + \kappa_T - \kappa_B + \kappa_L)} 
\ee
\be
\bar{S} = \frac{ S_{\widetilde H}^\cpv + S_{\widetilde t}^\cpv - S_{\widetilde b}^\cpv - S_{\widetilde \tau}^\cpv }{1 + \kappa_T - \kappa_B + \kappa_L } \;. \label{eq:Sbar}
\ee
\end{subequations}
Although the expressions in Eq.~\eqref{eq:Hdiff} are identical to that in the established literature~\cite{Huet:1995sh, Lee:2004we}, the form of Eqs.~\eqref{eq:bar} is dramatically different.  We note that there is no dependence on the first/second generation quark sector, owing to the fact that they do not participate in the dynamics which determines $n_L$.  

To solve Eq.~\eqref{eq:Hdiff} analytically, we follow Ref.~\cite{Huet:1995sh} making the approximations (a) that the true spatial dependence of the chiral relaxation rates may be replaced by a step-function, so that we may write $\bar{\Gamma}(z) = \bar{\Gamma} \, \theta(z)$; and (b) that $\bar{S}(z) \simeq 0$ for $z < -L_w/2$.
For the symmetric phase, where $z < -L_w/2$, we obtain
\be
H = \mathcal{A} \, e^{v_w z/\bar{D}} \;, \label{eq:Hsol}
\ee
where 
\be
\mathcal{A} = \int^\infty_0 dy \: \bar{S}(y) \frac{e^{-\gamma_+ y}}{\bar{D} \gamma_+} + \int^0_{-Lw/2} dy \: \bar{S}(y) \left[ \frac{\gamma_-}{v_w \gamma_+} + \frac{e^{-v_w y/\bar{D}}}{v_w} \right]\,. \label{eq:Acoeff}
\ee
Furthermore, we have defined
\be
\gamma_\pm \equiv \frac{1}{2 \bar{D}} \left[ v_w \pm \sqrt{ v_w^2 + 4 \bar{\Gamma} \bar{D} } \right] \;. \label{eq:gammapm}
\ee
We reiterate that although the form of Eqns.~(\ref{eq:Hsol}-\ref{eq:gammapm}) is similar to that in previous work~\cite{Huet:1995sh}, our results for $\bar{D}$, $\bar{\Gamma}$, and $\bar{S}$ are different, due to the modified structure of the Boltzmann equations in the presence of fast bottom and tau Yukawa rates.  

We now ask: was it safe to neglect $\Delta L$ in solving for $H$?  Substituting our solution for $H$ into Eq.~\eqref{eq:DeltaLR}, we find  that $\Delta L/H \propto 1/D_R$, in the limit that $D_R \to \infty$.  In short, in the physical limit where large RH lepton diffusion has the biggest impact upon $n_L$, our solution is most accurate.  There may be situations in general in which the impact of non-local corrections is not suppressed; we show how the Boltzmann equations may be solved in this case in future work \cite{chung}.

\section{Lepton-mediated electroweak baryogenesis: Numerical Results}
\label{sec:numerical}

We now consider an MSSM scenario that illustrates some of the novel
features discussed in Sec.~\ref{sec:analytic}.  As we will see, the
picture here is that the BAU is induced predominantly by leptonic
left-handed charge: hence, lepton-mediated.  The key parameters that
govern the behavior of this scenario are (i) $\tan\beta \gtrsim 20$
and pseudoscalar Higgs mass (at zero temperature) $m_A \lesssim 500$
GeV, ensuring $\tau_{y\tau}, \, \tau_{y\tau} \ll \tau_{\rm diff}$, and (ii) right-handed top and bottom squarks with approximately equal mass, thereby suppressing the quark contribution to $n_L$.  Here, we take both squarks to be light, with $\mathcal{O}(100 \; \textrm{GeV})$ masses, since a strong first order phase transition requires a light top squark.  

Although we work within the context of the MSSM, many of our conclusions are much more general.  In EWB scenarios beyond the MSSM, light squarks are not required for a strong first order phase transition (see e.g.~Refs.~\cite{Huber:2000mg,Menon:2004wv,Huber:2006wf}).  Even if the squarks are very heavy, EWB is still mediated by leptons occurs as long as the previous two conditions are met.

In this section, we first summarize the parameters of the lepton-mediated EWB scenario.  Next, we compute the bottom and tau Yukawa interaction rates $\Gamma_{yb}$ and $\Gamma_{y\tau}$, showing for what regions of parameter space they are fast compared to $\tau_{\textrm{diff}}$.  Last, we numerically solve the system of Boltzmann equations (\ref{eq:newbe2}-\ref{eq:lepeqn}) and compute the left-handed fermion charge density $n_L$ that generates $n_B/s$.  
Our main result is Fig.~\ref{fig:ewl}: it illustrates how $n_L$ arises from leptons instead of quarks, how our analytic and numerical results agree, and how this scenario differs dramatically from previous work neglecting $\Gamma_{yb}$ and $\Gamma_{y\tau}$.

\subsection{Input parameters}

\begin{table}[t]
\begin{tabular}{|c|c|c|c|c|c|c|c|c|c|c|}
\hline
$\mu$ & $120 \; \textrm{GeV}$ & & $M_T^2$ & $\; -(60 \; \textrm{GeV})^2$ & & $T$ & $\; 100 \; \textrm{GeV}$ & & $D_Q$ & $\; 6/T $\\
$M_1$ & $120 \; \textrm{GeV}$ & & $M_B^2$ & $(100 \; \textrm{GeV})^2$  & & $\; v(T)\, $ & $\; 125 \; \textrm{GeV}$ & & $\; D_H,\, D_L$ & $\; 100/T $\\
$M_2$ & $\; 250 \; \textrm{GeV}$ & & $\; M_R^2\,$ & $(300 \; \textrm{GeV})^2$  & & $\Delta\beta$ & $\; 0.015$ & & $D_R$ & $\; 380/T $ \\
$\; \tan\beta\,$ & $ 20   $ & &$m_A$ & $150 \; \textrm{GeV}$  & & $v_w$ & $\; 0.05 $ & & $L_w$ & $\; 25/T $ \\
\hline
\end{tabular}
\caption{Important parameters for lepton-mediated EWB scenario. \label{tab:ewl}}
\end{table}

The computation of $n_B/s$ relies upon many numerical inputs, some
described here and others described in the Appendix.  We have
evaluated the masses of particles during the EWPT assuming that
electroweak symmetry is unbroken.  This approximation is motivated by
the fact that most of the charge transport dynamics take place
outside the bubble in the region of unbroken symmetry.  These masses
receive contributions from the mass parameters in Table~\ref{tab:ewl}
and from finite temperature corrections, listed in the Appendix.  The
right-handed stop, sbottom, and stau SUSY-breaking mass-squared
parameters are $M^2_T$, $M^2_B$, and $M^2_R$, respectively.  The RH
stop is required to be light to achieve a strong first order phase
transition~\cite{Carena:2008vj}; taking the RH sbottom and stau to be
light as well ensures that the quark contribution to $n_L$ is
suppressed, while the lepton contribution is enhanced, in accord with
Eq.~\eqref{eq:lhcharge}.  We take all other squark and slepton
$($mass$)^2$ parameters to be 10 TeV.

For Higgs bosons, the story is more complicated.  Again, we study the degrees of freedom assuming unbroken electroweak symmetry.  
The mass term in the Lagrangian is
\be
\mathcal{L} \supset - \left(H_u^{+\dagger}, H_d^- \right) \left( \ba{cc} m_{u}^2 + |\mu|^2 + \delta_u  & b \\ b &  m_{d}^2 + |\mu|^2 + \delta_d \ea \right) \left( \ba{c} H_u^+ \\ H_d^{-\dagger} \ea \right) \;, \label{eq:Hmass}
\ee
and the same for $(H_u^0, H_d^{0\dagger})$ but with $b \to - b$.  The finite temperature corrections that restore electroweak symmetry are given by (see Table~\ref{tbl:thmass})
\begin{subequations}
\bea
\delta_u &=& \left( \, \frac{3}{8} \, g_2^2 + \frac{1}{8} \, g_1^2 + \frac{1}{2} \, y_t^2 \, \right) T^2 \\
\delta_d &=& \left( \, \frac{3}{8} \, g_2^2 + \frac{1}{8} \, g_1^2 + \frac{1}{2} \, y_b^2 + \frac{1}{6} \, y_\tau^2 \, \right) T^2 \; . 
\eea
\end{subequations}
(In the high $T$ limit, there are no off-diagonal thermal corrections, since these corrections are proportional to dimensionful parameters.)
We can re-express this mass matrix using the minimization conditions for electroweak symmetry breaking at $T=0$~\cite{Martin:1997ns}:
\begin{subequations}
\label{eq:mincon}
\bea
m_{u}^2 + |\mu|^2 &=& m_A^2 \cos^2 \beta + \frac{1}{2} m_Z^2 \cos 2\beta  \simeq - \frac{1}{2} \, m_Z^2 \\
m_{d}^2 + |\mu|^2 &=& m_A^2 \sin^2 \beta - \frac{1}{2} m_Z^2 \cos 2\beta  \simeq m_A^2 + \frac{1}{2} \, m_Z^2 \\
b &=& m_A^2 \sin\beta_0 \cos\beta \simeq 0\;,
\eea
\end{subequations}
where $m_Z$ and $m_A$ are the $Z$ and pseudoscalar Higgs boson masses at $T=0$.  The approximations in Eqs.~\eqref{eq:mincon} follow assuming $\tan\beta \gg 1$.  Therefore, in this limit, the Higgs boson mass matrix \eqref{eq:Hmass} is diagonal, with eigenvalues
\begin{subequations}
\label{eq:mH}
\bea
m_{H_u}^2 &=&  - \frac{1}{2} \, m_Z^2 + \delta_u \\
m_{H_d}^2 &=&  m_A^2 + \frac{1}{2} \, m_Z^2  + \delta_d  \;.
\eea
\end{subequations}
These are the Higgs boson masses during the EWPT.
We note that if $m_A \sim \mathcal{O}(100\; \textrm{GeV})$, then $H_d$ is also light.

The top, bottom, and tau Yukawa interactions are proportional to the corresponding Yukawa couplings.
At tree-level, these couplings are determined by
\be
\label{eq:ytbtau}
y_\tau = \frac{m_\tau}{v \, \cos\beta} \; , \qquad y_b = \frac{m_b}{v \, \cos\beta} \; , \qquad y_t = \frac{m_t}{v \, \sin\beta} \;,
\ee
where $v \simeq 174$ GeV is the Higgs vev at $T=0$.
However, quantum corrections lead to two complications. First, we include the QCD (QED) running of $y_b$ ($y_\tau$) from the scale where its mass is measured to the electroweak scale $m_Z$; this reduces $y_b$ by a factor $\eta_b \simeq 1.4$ and has negligible impact on $y_\tau$~\cite{Hall:1993gn}.  Second, we allow for the possibility that $y_{b,\tau}$ is smaller than expected at tree-level, due to $m_{b,\tau}$ receiving large one-loop corrections enhanced by $\tan\beta$, denoted as $\delta_b$ and $\delta_\tau$, for which we include only the dominant contributions~\cite{Hall:1993gn,masscorrections}.  Including both of these effects, the Yukawa couplings evaluated at the electroweak scale $m_Z$ are
\be
y_\tau(m_Z) = \frac{m_\tau}{v \, \cos\beta \, ( 1+ \delta_\tau \tan\beta)}
\; , \quad y_b(m_Z) = \frac{m_b/\eta_b}{v \, \cos\beta \, (1+ \delta_b
  \tan\beta)} \;.
\label{eq:runningandtanbeta}
\ee
For parameters given in Table \ref{tab:ewl}, we find $y_b(m_Z) \simeq 0.33$ and $y_\tau(m_Z) \simeq 0.20$.

The diffusion constants $D_i$ have been computed in Ref.~\cite{Joyce:1994bi,Joyce:1994zn}; the fact that $D_R \gg D_L$ enhances the left-handed lepton charge, as discussed in Sec.~\ref{sec:analytic}.  The bubble wall velocity $v_w$, thickness $L_w$, profile
parameters $\Delta\beta$ and $v(T)$ describe the dynamics of the
expanding bubbles during the EWPT, at temperature $T$~\cite{bubbles}.  
The spacetime-dependent vevs are approximated by
\bea
v(z) &\simeq& \frac{1}{2} \, v(T) \left[1 - \tanh\left(- \frac{3z}{L_w} \right) \right] \\
\beta(z) &\simeq& \beta(T) - \frac{1}{2} \, \Delta \beta \left[1 + \tanh\left(- \frac{3z}{L_w} \right) \right] \; ,
\eea
following Ref.~\cite{Carena:2000id}.

We consider a CP-violating source $S_{\widetilde H}^\cpv$
arising solely from Higgsino-Bino mixing, enhanced for $\mu = M_1$, and calculated following
Refs.~\cite{Riotto:1998zb,Lee:2004we}; the relevant CP-violating phase
is $\phi_1 \equiv \arg(M_1 \mu)$, and the Higgsino and Bino thermal widths are $\Gamma_{\widetilde H} \simeq 0.025 \, T$ and $\Gamma_{\widetilde B} \simeq 2\times 10^{-4} \, T$ (assuming the Bino width is dominated by quark-squark loops)~\cite{Elmfors:1998hh}.  Numerically, we find
\be
S^\cpv_{\widetilde H} \; \simeq \; - \, 0.24 \; \textrm{GeV} \; \times \beta^\prime(z) \, v(z)^2 \, \sin \phi_{1}  \;. 
\ee
This ``Bino-driven'' CP-violating source requires $|\phi_1| \gtrsim 1/20$, which is compatible with EDM searches provided we allow for $\arg(M_2 \mu) \ne \arg(M_1 \mu)$~\cite{Li:2008ez}.
The magnitude of $S_{\widetilde H}^\cpv$ --- and thus $n_B/s$ --- is proportional to $\Delta\beta$,
which itself goes as $\Delta\beta \, \propto \, 1/m_A^2$.  
Therefore, within this computation, 
viable EWB requires $m_A$ to be sufficiently light\footnote{The computation of the CP-violating source is the subject of ongoing scrutiny.  In other computations, there exist contributions to $S^\cpv_{\widetilde H}$ that are not suppressed as $m_A \to \infty$~\cite{Carena:2008vj}.}; in the next section,
we will see that light $m_A$ also increases the size of $\Gamma_{yb}$
and $\Gamma_{y\tau}$.   

The CP-conserving relaxation rates wash-out CP-violating asymmetries within the broken phase.  Computed following Ref.~\cite{Lee:2004we}, these rates are
\begin{align}
\Gamma_h(z) &\simeq 3.8 \times 10^{-3} \; \textrm{GeV}^{-1} \times v(z)^2 \\
\Gamma_{mt}(z) &\simeq 3.0 \times 10^{-3} \; \textrm{GeV}^{-1}  \times v(z)^2 \sin^2\beta(z) \\
\Gamma_{mb}(z) &\simeq \left(\frac{y_b}{y_t}\right)^2 \, \cot^2\beta(z) \: \Gamma_{mt}(z) \;.
\end{align}
We neglect additional CP-violating relaxation rates
from squarks, (s)leptons, and Higgs scalars.  

\subsection{Determination of bottom/tau Yukawa rates}

The leading contributions to $\Gamma_{yb}$ and $\Gamma_{y\tau}$ arise from absorption and decay processes in the thermal plasma, shown in Fig.~\ref{fig:feyn}a.  In the lepton-mediated scenario, the dominant processes are
\be
H_d \longleftrightarrow \, \bar{q}_L \, b_R \, , \; \bar{\ell}_L \, \tau_R \, , \qquad \qquad \widetilde{\tau}_R \longleftrightarrow \widetilde{H} \, \ell_L \; , \label{eq:decays}
\ee
where $q_L \!=\! t_L, \, b_L$, and $\ell_L\!=\! \tau_L, \, \nu_\tau$.  
(Processes with left-handed squarks are suppressed since $M_Q^2 = (10$ TeV$)^2$; in addition, the decays $\widetilde{b}_R \longleftrightarrow \widetilde{H} \, q_L$ are kinematically forbidden.)
We compute these thermally-averaged decay rates following Ref.~\cite{Cirigliano:2006wh}; they are
\begin{subequations}
\bea
\Gamma_{yb} &=& \frac{12 \, N_c \, y_b^2}{T^2} \: \mathcal{I}_{F}(m_{q_L}, \, m_{H_d}, \, m_{b_R}) \;, \\
\Gamma_{y\tau} &=& \frac{12 \, y_\tau^2}{T^2} \: \left( \, \mathcal{I}_{F}(m_{\ell_L}, \, m_{H_d}, \, m_{\tau_R}) + \mathcal{I}_{F}(m_{\widetilde H}, \, m_{\widetilde\tau_R}, \, m_{\ell_L}) \frac{}{} \right) \;.
\eea
\end{subequations}
We refer the reader to Ref.~\cite{Cirigliano:2006wh} for the general form of $\mathcal{I}_F$, which is the form used in our numerical analysis.  For the case of a scalar decaying into two fermions, $\phi \leftrightarrow \psi_1 \, \bar{\psi}_2$, it is approximately given by
\be
\mathcal{I}_F(m_1, \, m_\phi, \, m_2) \simeq \frac{T^3}{4} \, \left( \frac{m_\phi}{2\pi T} \right)^{5/2} \, \left[1 - \left(\frac{m_1+m_2}{m_\phi} \right)^2 \, \right] \, e^{-m_\phi/T}  \;.\label{eq:IFsimp}
\ee
This simplied form for $\mathcal{I}_F$ is obtained by assuming $m_\phi > m_1 + m_2 \gg |m_1 - m_2|$, and taking Maxwell-Boltzmann statistics for these particles; it is valid at the $\mathcal{O}(25\%)$ level for $m_\phi \gtrsim 2(m_1 + m_2)$.

\begin{figure*}
\includegraphics[scale=0.8]{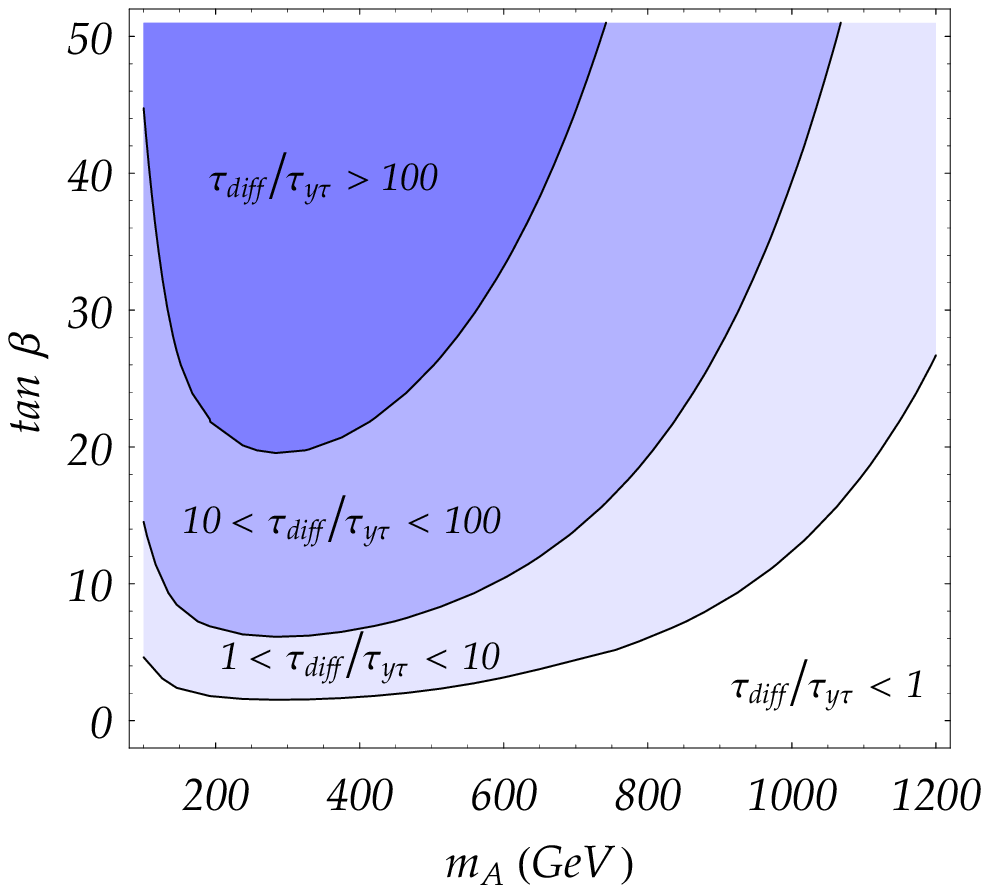} \includegraphics[scale=0.8]{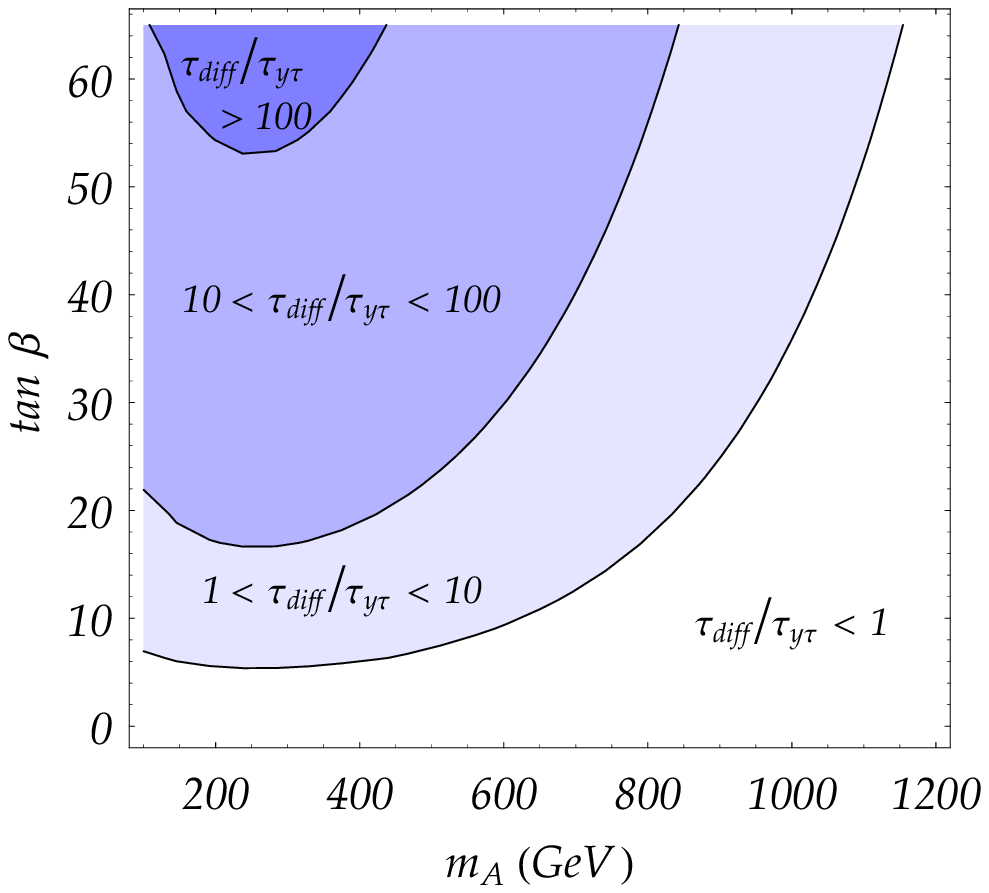}
\caption{Contour plot of $\tau_{\rm diff}/\tau_{yb}$ (left) and $\tau_{\rm diff}/\tau_{y\tau}$ (right) in $\tan\beta$-$m_A$ parameter space.  Values of $\tau_{yb}$ and $\tau_{y\tau}$ include contributions from $H_d \leftrightarrow q_L b_R$ and $H_d \leftrightarrow \ell_L \tau_R$ only.  Tau Yukawa chemical equilibrium is maintained for $\tau_{\rm diff}/\tau_{y\tau} \gtrsim 10$; similarly for the bottom Yukawa.}
\label{fig:contour}
\end{figure*}

The key parameters that govern $\Gamma_{y\tau}$ and $\Gamma_{yb}$ are $\tan\beta$ (since the Yukawa couplings $y_{b,\tau} \propto \tan\beta$ for $\tan\beta \gg 1$) and the masses of the particles in the decays in Eq.~\eqref{eq:decays}.  From Eq.~\eqref{eq:IFsimp}, we see that when the decaying scalars $H_d$ and $\widetilde\tau_R$ are light compared to $T$, the rates $\Gamma_{y\tau}$ and $\Gamma_{yb}$ are largest.  In particular, as shown in Eq.~\eqref{eq:mH}, the mass of $H_d$ during the EWPT is related to the masses of the $Z$ and pseudoscalar Higgs at $T=0$.  Not only does light $m_A$ enhance the CP-violating source, it enhances $\Gamma_{yb}$ and $\Gamma_{y\tau}$.  We illustrate this fact in Fig.~\ref{fig:contour}: here, we show the time scales for these rates, $\tau_{yb} \equiv \Gamma_{yb}^{-1}$ (left) and $\tau_{y\tau} \equiv \Gamma_{y\tau}^{-1}$ (right), compared to the diffusion time scale $\tau_{\textrm{diff}} \equiv \bar{D}/v_w^2$, as a function of $\tan\beta$ and $m_A$.  We only include contributions to these rates from $H_d$ decay.  Bottom Yukawa interactions are in chemical equilibrium for $\tau_{\textrm{diff}}/\tau_{yb} \gtrsim 10$, when $\tan\beta \gtrsim 5$ and $m_A \lesssim 600$ GeV; similarly, tau Yukawa interactions are in chemical equilibrium for $\tau_{\textrm{diff}}/\tau_{y\tau} \gtrsim 10$, when $\tan\beta \gtrsim 20$ and $m_A \lesssim 500$ GeV.  For the parameters chosen earlier, we have
\be
\Gamma_{yb} \simeq 0.28 \; \textrm{GeV} \, , \qquad \Gamma_{y\tau} \simeq 0.12 \; \textrm{GeV} \, , \qquad \Gamma_{yt} \simeq 2.4 \; \textrm{GeV} \; ,
\ee
with $\Gamma_{yt}$ computed in Ref.~\cite{Cirigliano:2006wh}; for comparison, $\tau_{\textrm{diff}}^{-1} \simeq 0.005\:$ GeV.

Scattering contributions, shown in Fig.~\ref{fig:feyn}b, also contribute to bottom and tau Yukawa rates.  They are suppressed in comparison by $\alpha_s$ or $\alpha_w$, but become the dominant contribution when absorption and decay are kinematically forbidden.  For completeness, we also note another class of interactions mediated by third generation Yukawa couplings: F-term-induced four-scalar interactions, shown in Fig.~\ref{fig:feyn}c.  However, one can show that if all $\Gamma_{yi}$ (for $i = t,b,\tau$) interactions are in chemical equilibrium, then chemical equilibrium is satisfied for these four-scalar interactions as well.  
In the present work, we consider only the contributions to $\Gamma_{yb}$ and $\Gamma_{y\tau}$ that arise from absorption and decay processes.  Therefore, our calculation for these rates is a lower bound; the extent to which they lead to chemical equilibrium can only be enhanced by the inclusion of scattering processes in Figs.~\ref{fig:feyn}b,c.

\subsection{Numerical Solution to Boltzmann Equations}
\begin{figure*}[t]
\includegraphics[scale=.8]{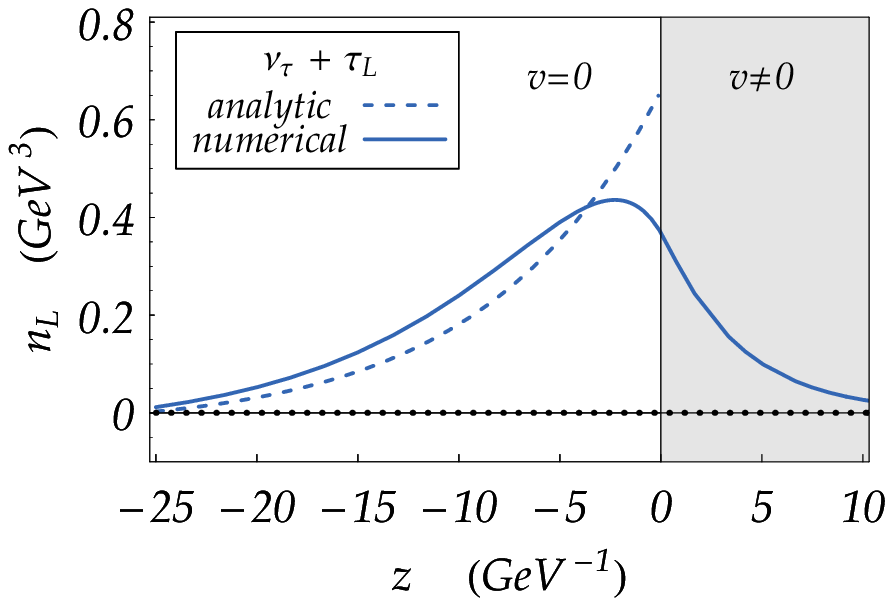} \includegraphics[scale=.8]{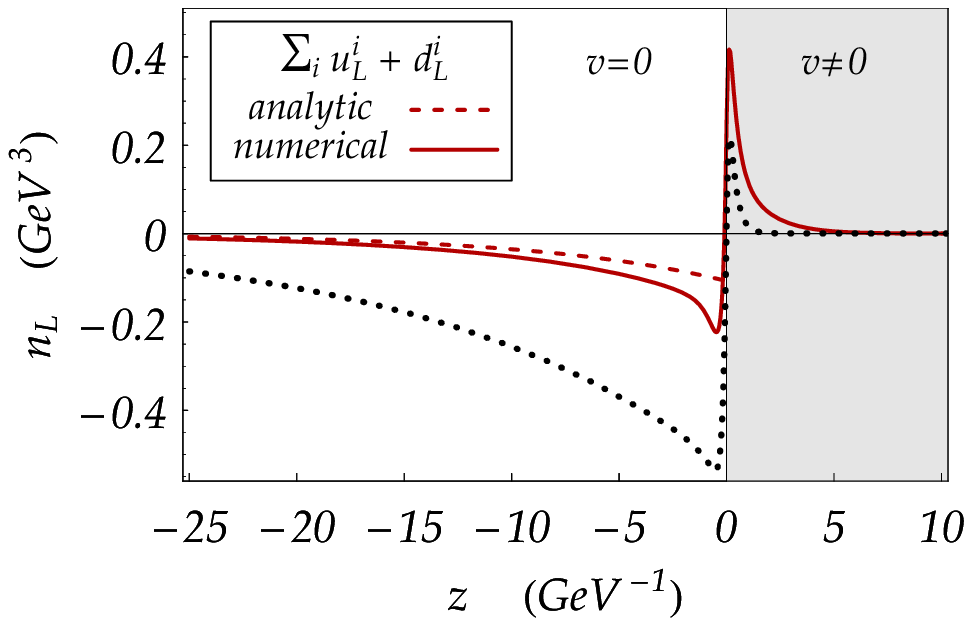}
\caption{Left-handed charge densities for leptons (left panel) and quarks (right panel) that generate $n_B/s$, for lepton-mediated scenario.  Solid (dashed) curves are numerical (analytic) results, as function of distance $z$ from bubble wall.  Shaded region denotes broken electroweak symmetry.  Dotted curves are numerical results obtained neglecting tau/bottom Yukawa interactions.  The effect of these interactions is to suppress LH quark charge, while enhancing LH lepton charge, thereby flipping the sign of $n_L$ and $n_B/s$ compared to previous computations.  }
\label{fig:ewl}
\end{figure*}
We now solve the Boltzmann equations (\ref{eq:newbe2}-\ref{eq:lepeqn}) numerically for the lepton-mediated EWB scenario.  
In Fig.~\ref{fig:ewl}, we show the left-handed fermion charge density $n_L$ that arises from leptons (left) and quarks (right), for maximal CP-violating phase $\phi_1 = - \pi/2$.  Our numerical results are shown by the solid curves, plotted as a function of the distance $z$ to the moving bubble wall.  The region of broken electroweak symmetry (denoted $v\ne 0$) is for $z>0$, while unbroken symmetry is for $z<0$ (denoted $v=0$).  As promised, $n_L$ is predominantly leptonic, while the quark contribution is suppressed.  The dashed curves are our analytic results (plotted only for $z<0$), obtained in Sec.~\ref{sec:analytic}.  Our analytic and numerical results are in good agreement. However, close to the bubble wall,  there is some disagreement between numerical and analytic lepton charge densities.  For $|z| \lesssim \sqrt{\bar D \, \tau_{y\tau} } \simeq 2$ GeV, the lepton density has not had enough time to reach chemical equilibrium; here, our analytic approximation is breaking down, as discussed in Sec.~\ref{sec:simpboltzmann}.  

The resulting baryon asymmetry is
\be
n_B/s \; \simeq \; \left\{ \ba{cc} 8 \times \sin\phi_1 \, \left(n_B/s\right)_{\textrm{CMB}} & \qquad \textrm{Bottom/tau Yukawas included} \\
-14 \times \sin\phi_1 \, \left(n_B/s\right)_{\textrm{CMB}} & \qquad \textrm{Bottom/tau Yukawas neglected} \ea \right. \; ,
\ee
where 
$(n_B/s)_{\textrm{CMB}} =
8.84\times 10^{-11}$ is the central value obtained from the
CMB~\cite{Dunkley:2008ie}.  That is, a CP-violating phase $\sin\phi_1 \simeq 1/8$ is required to give the observed BAU in this scenario.  However, if we had neglected bottom and tau Yukawa interactions, the picture completely flips, as shown by the dotted curves in Fig.~\ref{fig:ewl}.  In this case, the lepton charge vanishes (left panel) and the quark charge becomes dominant (right panel), thereby flipping
the sign of the required CP-violating phase: $\sin\phi_1 \simeq -1/14$.  In the lepton-mediated scenario, the impact of bottom and tau Yukawa interactions is dramatic.  If electric dipole moment searches uncover new CP-violating phases, such as $\phi_1$, the inclusion $\Gamma_{yb}$ and $\Gamma_{y\tau}$ will clearly be essential in testing the consistency of supersymmetric EWB scenarios.

\section{Conclusion}
\label{sec:conclude}

In this work, we have studied how the transport dynamics involving the tau and bottom Yukawa interactions impact the generation of left-handed fermionic charge $n_L$ that biases baryon number production via electroweak sphalerons.
Previous work has neglected these interactions.  However, we showed that these interactions are not negligible and can have a dramatic impact upon $n_L$ and $n_B/s$.  When tau and bottom Yukawa interactions are in chemical equilibrium, the following effects occur:
\begin{itemize}
\item Significant third generation lepton charge is generated.  In contrast, without $\Gamma_{y\tau}$, no left-handed lepton asymmetry is generated.  We showed how this charge is enhanced by efficient diffusion of right-handed leptons.  
\item No first and second left-handed quark charge is generated, and
strong sphaleron processes are unimportant.  In contrast, without
$\Gamma_{yb}$, a significant fraction of left-handed charge comes from
first and second generation quarks, and strong sphaleron processes become
important.
\item Third generation left-handed quark charge is suppressed if the
  right-handed top and bottom squarks have equal mass, or if their masses are large compared to the temperature.  In contrast,
  without $\Gamma_{yb}$, this suppression does not occur.
\end{itemize}
In light of these differences, the inclusion of $\Gamma_{yb}$ and
$\Gamma_{y\tau}$ can have a strong impact on the computation of the
baryon asymmetry.

To verify our analytic conclusions, we calculated the rates $\Gamma_{yb}$ and
$\Gamma_{y\tau}$, and solved the full system of Boltzmann equations
numerically.  
We considered a ``lepton-mediated electroweak baryogenesis scenario'', where all three right-handed third
generation scalars have $\mathcal{O}(100 \; $GeV$)$ masses, and all
other squarks and sleptons have $\mathcal{O}($TeV$)$ masses.  Here, we
showed that all three left-handed quark densities are suppressed, and
the baryon asymmetry is induced primarily from left-handed lepton
charge.  The CP-violating phase required for the observed baryon
asymmetry is positive.  In contrast, if we had neglected $\Gamma_{yb}$
and $\Gamma_{y\tau}$, we would have arrived at a completely different
picture: the baryon asymmetry would have arisen from left-handed quark
density, and the required CP-violating phase would have been negative.
The baryon asymmetry in the this scenario is strongly impacted by the
presence of $\Gamma_{yb}$ and $\Gamma_{y\tau}$.  Although we have focused on a lepton-mediated scenario, the leptonic component of $n_L$ can be important even when the quark component is not suppressed, enhancing or suppressing the total BAU.
We conclude,
therefore, that bottom and tau Yukawa rates are relevant and must be
included in supersymmetric electroweak baryogenesis computations, both
in the MSSM and beyond.

\begin{acknowledgments}

This work was supported in part by Department of Energy contracts
DE-FG02-08ER41531 and DE-FG02-95ER40896, and the Wisconsin Alumni Research
Foundation.

\end{acknowledgments}

\begin{appendix}

\section{Numerical inputs}

Here we summarize other numerical inputs needed to compute $n_B/s$.
In Table \ref{tbl:thmass}, we list various finite temperature mass corrections that relevant for EWB.  The numerical values of gauge couplings used are $g_1=0.357, g_2=0.652, g_3=1.23$ \cite{Yao:2006px}.  We note that $\widetilde H_{u}, \, \widetilde{H}_d$ have different finite temperature contributions; here, we treat these degrees of freedom as a single Dirac fermion with plasma mass equal to the average of these contributions.

\begin{table}
\begin{tabular}{|l||l|l|l|}
\hline 
 Particle & $\; \delta m_{{\textrm{SM}}}^{2}/T^{2}\quad (A)$  & $\; \delta m_{\textrm{SUSY}}^{2}/T^{2} \quad (B)$  & $\; \delta m_{\textrm{SUSY}}^{2}/T^{2} \quad (C)$ \tabularnewline
\hline 
$\;q_{L}$  & $\frac{1}{6}g_{3}^{2}+\frac{3}{32}g_{2}^{2}+\frac{1}{288}g_{1}^{2}+\frac{1}{16}y_{t}^{2}+\frac{1}{16}y_{b}^{2}$  & $+\frac{1}{16}y_{t}^{2}$  & $+\frac{1}{16}y_{b}^{2}$ \tabularnewline
$\;t_{R}$  & $\frac{1}{6}g_{3}^{2}+\frac{1}{18}g_{1}^{2}+\frac{1}{8}y_{t}^{2}$  & $+\frac{1}{18}g_{1}^{2}$ &  \tabularnewline
$\;b_{R}$  & $\frac{1}{6}g_{3}^{2}+\frac{1}{72}g_{1}^{2}+\frac{1}{8}y_{b}^{2}$  &  & $+\frac{1}{72}g_{1}^{2}$ \tabularnewline
$\;\ell_L$  & $\frac{3}{32}g_{2}^{2}+\frac{1}{32}g_{1}^{2}+\frac{1}{16}y_{\tau}^{2}$  &  & $+\frac{1}{16}y_{\tau}^{2}$ \tabularnewline
$\;\tau_{R}$  & $\frac{1}{8}g_{1}^{2}+\frac{1}{8}y_{\tau}^{2}$  &  & $+\frac{1}{8}g_{1}^{2}$ \tabularnewline
$\;\widetilde{H}_{u}$  &  & $\frac{3}{16}g_{2}^{2}+\frac{1}{16}g_{1}^{2}+\frac{3}{16}y_{t}^{2}$  &   \tabularnewline
$\;\widetilde{H}_{d}$  &  & $\frac{3}{16}g_{2}^{2}+\frac{1}{16}g_{1}^{2}$  & $+\frac{3}{16}y_{b}^{2}+\frac{1}{16}y_{\tau}^{2}$ \tabularnewline
$\;\widetilde{W}$  &  & $\frac{3}{8}g_{2}^{2}$  &  \tabularnewline
$\;\widetilde{B}$  &  & $\frac{5}{12}g_{2}^{2}$  & $+\frac{2}{12}g_{2}^{2}$ \tabularnewline
$\;\widetilde{t}_{R}$  &  & $\frac{4}{9}g_{3}^{2}+\frac{1}{3}g_{1}^{2}+\frac{1}{3}y_{t}^{2}$  & $-\frac{1}{9}g_{1}^{2}$ \tabularnewline
$\;\widetilde{b}_{R}$  &  &  & $\frac{4}{9}g_{3}^{2}+\frac{1}{18}g_{1}^{2}+\frac{1}{3}y_{b}^{2}$ \tabularnewline
$\;\widetilde{\tau}_{R}$  &  &  & $\frac{1}{2}g_{1}^{2}+\frac{1}{3}y_{\tau}^{2}$ \tabularnewline
$\;H_{u}$  & $\frac{3}{16}g_{2}^{2}+\frac{1}{16}g_{1}^{2}+\frac{1}{4}y_{t}^{2}$  & $+\frac{3}{16}g_{2}^{2}-\frac{1}{48}g_{1}^{2}+\frac{1}{4}y_{t}^{2}$  & $+\frac{1}{12}g_{1}^{2}$  \tabularnewline
$\;H_{d}$  & $\frac{3}{16}g_{2}^{2}+\frac{1}{16}g_{1}^{2}+\frac{1}{4}y_{b}^{2}+\frac{1}{12}y_{\tau}^{2}$  & $+\frac{3}{16}g_{2}^{2}+\frac{7}{48}g_{1}^{2}$  & $-\frac{1}{12}g_{1}^{2}+\frac{1}{4}y_{b}^{2}+\frac{1}{12}y_{\tau}^{2}$ \tabularnewline
\hline
\end{tabular}
\caption{\label{tbl:thmass} Thermal masses $\delta m^2$ for active particles in the plasma during EWPT.  The different contributions arise from thermal loops involving: (A) SM fermions and gauge bosons only; (B) Higgsinos, Winos, Binos, and RH stops; and (C) RH sbottoms and RH staus.  In the lepton-mediated scenario, all three contributions must summed.}
\end{table}

There are additional parameters of minor relevance to the EWB computation.  We choose third generation SUSY-breaking triscalar parameters to be $(y_i \, A)$, for $i=t,b,\tau$, with a common $A=7$ TeV.  In addition, we choose gluino mass parameter $M_3 = 500$ GeV.  These parameters are relevant for computing $\delta_{b,\tau}$ in Eq.~\eqref{eq:ytbtau}.  In addition, with these parameters, we can compute the $T=0$ squark and slepton spectrum; the lightest, mostly RH, third generation scalars have masses
\be
m_{\widetilde t_1} \simeq 103 \; \textrm{GeV} \, , \qquad  m_{\widetilde b_1} \simeq 93 \; \textrm{GeV} \, , \qquad m_{\widetilde \tau_1} \simeq 297 \; \textrm{GeV}\; .
\ee
All other squarks and sleptons have $\mathcal{O}(10 \; \textrm{TeV})$ masses.  The computation of $n_L$ is effectively insensitive to these heavy masses, since these particles have decoupled from the plasma before the EWPT.

Lastly, we describe how to obtain $n_B/s$ after solving the system of transport equations for $n_{L}(z)$ --- either analytically or numerically, as described in the text.  The BAU is given by
\be
n_B = - \frac{3 \Gamma_{ws}}{2 D_Q \lambda_+} \int^{-L_w/2}_{-\infty} dz \; n_{L}(z) \, e^{-\lambda_- z} \;.
\ee
with
\be
\lambda_\pm = \frac{1}{2D_Q} \left( v_w \pm \sqrt{ v_w^2 + 4 D_Q \, \mathcal{R} } \right) \;,
\ee
with $\mathcal{R} \simeq 2 \times 10^{-3}$ GeV, the wash-out rate for inverse electroweak sphaleron transitions (see e.g.~Ref.~\cite{Lee:2004we}).  Baryon number induced by $n_L(z)$ is washed out for $z \gg |\lambda^{-1}_-| \simeq 30$ GeV$^{-1}$; this is another reason to neglect the non-local contributions, which contribute to $n_L$ far from the bubble wall ($z=0$).
Lastly, the entropy density is given by
\be
s = \frac{2 \pi^2}{45} \, g_{*S} \, T^3 \;,
\ee
where $g_{*S} \simeq 131$, for the parameters given.

\end{appendix}

\end{document}